\documentclass[final,5p,times]{elsarticle}

\usepackage{doi}
\usepackage{xurl}
\usepackage{microtype}
\usepackage{amsmath,amssymb}
\usepackage{graphicx}
\usepackage{booktabs}
\usepackage{tikz}
\usetikzlibrary{arrows.meta, positioning}
\usetikzlibrary{fit, calc}
\usepackage{float}
\usepackage{placeins}
\usepackage{pgfplots}
\usepgfplotslibrary{fillbetween}
\pgfplotsset{compat=1.18}
\usepackage{tabularx}
\usepackage{seqsplit}  
\usepackage{multirow}
\usepackage{siunitx}
\usepackage{hyperref}
\begin{document}

\begin{frontmatter}

\title{RepliCore: Reproducible Parallel Simulation under Asynchronous Browser Runtimes}

\author[inst1]{Anqing Chen\corref{cor1}}
\ead{caqing1@cdtu.edu.cn}

\cortext[cor1]{Corresponding author}

\address[inst1]{School of Computer Engineering, Chengdu Technological University, Chengdu, China}

\begin{abstract}

Browser-based simulations execute over asynchronous runtime mechanisms including event loops, rendering callbacks, and independently scheduled Web Workers, causing simulation progression to depend on runtime timing and callback scheduling behavior.

RepliCore addresses this problem by separating asynchronous runtime progression from externally observable logical-state visibility. Runtime-visible simulation states are exposed only after logical-state progression becomes externally stable, preventing asynchronous runtime activities from observing partially updated simulation progression. The framework prevents rendering callbacks and asynchronous runtime tasks from observing transient intermediate logical states during parallel progression. This organization maintains consistency between parallel and sequential execution.

Based on this model, we implement RepliCore, a browser-oriented deterministic parallel simulation framework for reproducible large-scale simulation under asynchronous browser runtimes. Experiments in real browser environments produce bitwise-identical outputs across varying worker configurations, scheduling conditions, and rendering frequencies while remaining practical for large-scale browser-oriented workloads. Additional ablation experiments show systematic state divergence after relaxing key execution constraints.

These results indicate that reproducible asynchronous parallel simulation can be achieved through controlled logical-state visibility stabilization without relying on execution replay or explicit schedule control.

\end{abstract}

\begin{keyword}
Browser-Based Simulation,
Deterministic Parallel Simulation,
Runtime Reproducibility,
Asynchronous Execution,
Web Workers,
Parallel Runtime Systems
\end{keyword}

\end{frontmatter}

\section{Introduction}

Browser-oriented simulations increasingly execute over asynchronous browser runtime environments composed of event loops, rendering callbacks, independently scheduled Web Workers, and concurrent runtime services~\cite{whatwg_event_loop}. Under these conditions, simulation progression may become sensitive to runtime timing behavior and asynchronous callback scheduling. Because Workers progress independently, partially updated logical states may become externally observable at different physical times across repeated executions.

This runtime variability may expose different logical-state trajectories across repeated executions of browser-oriented simulation systems.
Logical-state progression may become dependent on execution interleavings and scheduling behavior in addition to simulation inputs~\cite{gonnord2023survey}. As simulation scale increases, maintaining reproducible execution behavior across asynchronous browser runtimes becomes increasingly difficult~\cite{peng2011reproducible,collberg2016repeatability,antunes2024repro}.

Existing approaches to reproducible runtime execution often attempt to preserve consistent state progression through scheduling control, strict coordination, or execution replay. These assumptions become increasingly fragile in browser runtimes, where scheduling behavior is affected by independently progressing Workers and asynchronous runtime activities that are not directly controlled by the simulation itself. Rendering updates, timer callbacks, user interactions, and network events may all influence execution timing during simulation progression. For example, runtime callbacks may access partially updated partition states while updates to other partitions are still in progress. Meanwhile, rendering callbacks or asynchronous event handlers may observe partially updated simulation data before stable logical states become externally visible. Repeated executions may therefore expose different externally observable state trajectories even under identical simulation inputs and initial conditions.

Under these runtime conditions, reproducibility depends heavily on whether transient intermediate logical states become externally observable during parallel progression.

To address this issue, RepliCore constrains when updated logical states become externally observable during asynchronous runtime execution. Workers may complete local updates at different physical times, while rendering callbacks and asynchronous runtime tasks continue to execute concurrently. Without controlled visibility boundaries, partially updated logical states may become externally observable before global progression reaches a stable state. RepliCore therefore delays externally visible state exposure until logical-state progression becomes externally stable.

Under this organization, Worker scheduling order, rendering frequency, and execution interleavings no longer determine final simulation results.

We implement RepliCore as a browser-oriented deterministic parallel simulation framework and evaluate it under varying Worker configurations and asynchronous runtime conditions. The evaluation demonstrates reproducible execution behavior across changes in Worker scheduling, rendering frequency, and parallel execution scale while maintaining stable scalability for large simulation workloads. The results further indicate that reproducibility depends strongly on controlling externally observable logical-state exposure under asynchronous runtime conditions.

The main contributions of this paper are as follows:

\begin{itemize}

\item An analysis of how asynchronous browser runtime activities introduce nondeterministic logical-state visibility during parallel simulation execution;

\item A logical-state visibility stabilization model for deterministic progression under asynchronous browser runtime execution;

\item The design and implementation of RepliCore, a browser-oriented simulation framework that constrains externally observable logical-state visibility under asynchronous runtime execution;

\item Experimental evidence showing reproducible execution across varying Worker scheduling and rendering conditions, together with ablation results demonstrating accumulated divergence after relaxing logical-state visibility constraints.

\end{itemize}
\section{Related Work}

\subsection{Non-determinism and Reproducibility in Parallel Systems}

Maintaining consistent state progression remains a major challenge in asynchronous parallel and distributed runtime environments. Runtime scheduling variability, asynchronous callback execution, and timing-dependent state visibility may cause repeated executions to expose different logical-state trajectories even under identical inputs~\cite{bernstein1966analysis,herlihy2020art}.

Large-scale data-driven computing and simulation systems have increased the difficulty of maintaining reproducible execution. As simulation and data-processing pipelines have become increasingly parallel and distributed, reproducibility problems now affect validation, debugging, and scientific result verification simultaneously. Recent work in computational science and systems research has reported persistent reproducibility problems across software and simulation environments that affect the consistency and verification of scientific results~\cite{stodden2016Enhancing,peng2011reproducible,collberg2016repeatability}.

In HPC and large-scale systems, reproducibility becomes harder to maintain because execution depends on complex software stacks, parallel execution models, and heterogeneous hardware platforms~\cite{dongarra2019hpc,antunes2024repro,perumalla2006Parallel,costa2026SciRep}. These factors make end-to-end reproducibility difficult to maintain across different runtime environments.

Recent surveys have also summarized the relationship between parallel execution and deterministic behavior across different execution models~\cite{gonnord2023survey}. Earlier theoretical work analyzed several fundamental sources of non-determinism. For example, the Bernstein conditions~\cite{bernstein1966analysis} describe independence relations between concurrent tasks, and logical clocks~\cite{lamport1978time} establish causal ordering between distributed events. These models describe ordering and causality relations between concurrent events, but they do not specify how parallel state updates should be organized to preserve identical simulation results across different runtime schedules.

\vspace{0.5em}
\noindent
\textbf{Limitation.}
Existing approaches explain several sources of non-determinism, but they do not directly provide an execution model for deterministic parallel execution.

\subsection{Runtime Control and Reconstruction-Based Approaches}

Existing deterministic execution techniques mainly rely on either runtime ordering control or post hoc execution reconstruction.

\textbf{(1) Runtime ordering control.}
Execution-control approaches reduce scheduling variability through deterministic schedulers, constrained synchronization patterns, or restricted shared-memory semantics~\cite{bocchino2009DPJ,lee2006problem,aviram2010deterministic}. These systems typically assume that thread execution, synchronization behavior, and memory visibility can be coordinated sufficiently to preserve stable execution ordering across runs.

\textbf{(2) Execution reconstruction.}
Execution-reconstruction approaches reproduce prior executions by recording runtime traces and replaying observed execution sequences~\cite{leblanc1987debugging,netzer1993optimal,geels2007replay,veeraraghavan2012doubleplay}. Later systems reduce replay overhead through lightweight tracing, optimized logging, and production-oriented replay infrastructures~\cite{liu2011dthreads,fu2026multithreads,replay2017rr}. Browser-oriented replay systems have also been proposed, such as Mugshot for deterministic capture and replay of JavaScript applications~\cite{mickens2010mugshot}.

Both categories depend on the ability to observe, constrain, or reproduce runtime ordering behavior.

\vspace{0.5em}
\noindent
\textbf{Limitation.}
Browser runtimes introduce several additional sources of scheduling variability that are difficult to coordinate through conventional deterministic scheduling or replay mechanisms. Event callbacks may originate from independent task queues, rendering updates may be triggered asynchronously by the rendering pipeline, and Worker communication may interleave differently depending on runtime timing conditions. Callback dispatch timing may also vary across browser implementations, rendering load, and external event activity.

Consequently, the runtime ordering observed during one execution may differ across repeated runs even when the application-level logic remains unchanged. Recording and replaying all relevant asynchronous interactions can also introduce substantial tracing and runtime-coordination overhead, particularly when simulation progression interacts continuously with rendering callbacks, Worker messages, and external event delivery.

\subsection{Parallel Discrete Event Simulation (PDES)}

Parallel Discrete Event Simulation (PDES) is a widely used approach for large-scale parallel simulation in which events are processed concurrently under causal-ordering constraints~\cite{fujimoto1990parallel}.

Typical PDES mechanisms include:
\begin{itemize}
\item Conservative approaches (e.g., Chandy--Misra) enforce causal ordering by restricting the execution of unsafe events~\cite{chandy1979distributed}
\item Optimistic approaches (e.g., Time Warp) permit speculative execution and recover from causality violations through rollback mechanisms~\cite{jefferson1985virtual}
\end{itemize}

Recent PDES systems mainly target scalability and runtime performance in multi-core and distributed execution environments~\cite{jia2025scalePDES}.

Most PDES systems prioritize causal correctness and scalable event execution.
Whether identical simulation results are preserved across different thread schedules, partitioning strategies, or runtime timing conditions is typically left to the implementation.

Although deterministic PDES variants exist, they generally address reproducibility through event-order control and causality management within simulation engines. RepliCore instead focuses on deterministic logical-state visibility under browser-runtime execution, where rendering activity, event-loop progression, and asynchronous Worker coordination introduce additional sources of runtime variability.

\subsection{Comparison with BSP and Dataflow Models}

The proposed execution model resembles several existing parallel execution paradigms, including Bulk Synchronous Parallel (BSP), dataflow systems, and synchronous simulation frameworks. These systems commonly use logical-time progression, partitioned computation, and runtime-wide coordination during execution. The primary difference is that RepliCore focuses on deterministic observable-state construction under asynchronous browser runtime execution, rather than primarily targeting phase-oriented parallel coordination.

\textbf{Bulk Synchronous Parallel (BSP).}
BSP organizes computation into supersteps separated by synchronization points~\cite{valiant1990bridging,mcsherry2015timely}. These synchronization stages help structure communication and parallel execution, but reproducibility may still depend on implementation-level factors such as aggregation order, floating-point accumulation behavior, and timing-dependent state exposure. RepliCore instead focuses on maintaining stable logical-state progression under asynchronous browser runtime behavior involving rendering callbacks, event-loop scheduling, and independently progressing Workers.

\textbf{Dataflow and Functional Models.}
Dataflow and functional systems organize computation as dependency-driven operations, often represented through directed acyclic graphs or pure functions~\cite{dennis1974first,akkiraju2015dataflow}. These systems often produce deterministic behavior at the level of individual operations. However, they mainly focus on deterministic local computation and dependency propagation, rather than maintaining consistent system-wide results across different asynchronous execution conditions.

\textbf{Synchronous and Time-Stepped Simulation.}
Synchronous simulation frameworks advance system state in discrete time steps and typically rely on coordinated logical-time execution to maintain consistent simulation behavior~\cite{fujimoto2000parallel}. In practice, implementation-level nondeterminism may still arise from parallel execution details within a simulation step. RepliCore separates logical-step computation from runtime scheduling by executing each step against a committed logical snapshot and constructing the next globally visible state through deterministic aggregation.

\vspace{0.5em}
\noindent
\textbf{Key Distinction.}
The approaches above mainly coordinate parallel execution through synchronization, dependency management, or logical-time progression. RepliCore primarily targets deterministic logical-state progression under browser runtime execution, where rendering behavior, callback scheduling, and Worker timing may otherwise expose inconsistent intermediate states across runs.

Many existing systems maintain reproducibility through execution coordination, synchronization, or ordering mechanisms. RepliCore constrains state updates and aggregation behavior so that simulation results remain stable across different execution schedules in highly asynchronous environments such as modern Web systems.

\subsection{Web Asynchronous Execution Model}

Modern Web runtimes execute applications through an event-driven scheduling model centered around the browser event loop~\cite{whatwg_event_loop}. Application execution is not driven by a single sequential control flow, but instead progresses through asynchronously dispatched callbacks originating from multiple runtime subsystems.

In browser environments, callback execution may be triggered by independent task queues, rendering updates, user interaction events, network activity, and inter-Worker communication. Rendering progression and application computation are also only partially synchronized. For example, rendering callbacks such as \url{requestAnimationFrame}~\cite{gregory2018game} are coordinated by the rendering pipeline, while Worker execution proceeds independently from rendering activity. Message delivery between Workers and the main thread may therefore interleave differently depending on rendering load, event-loop timing, and callback scheduling conditions.

These runtime characteristics introduce timing-dependent differences in callback execution order, Worker completion timing, and state-update visibility. During parallel simulation execution, partially updated states may therefore become observable at different times across executions, causing logical state progression to depend indirectly on runtime scheduling behavior.

WebAssembly and WebGPU have enabled browsers to support increasingly complex general-purpose computation~\cite{haas2017wasm,jangda2019wasm,w3c:webgpu}. Most existing Web systems research primarily treats the browser as a portable execution substrate for compute acceleration, with greater emphasis on throughput and portability than runtime execution semantics.

\vspace{0.5em}
\noindent
\textbf{Limitation.}
Maintaining reproducible simulation progression under browser runtime conditions remains difficult because logical updates are often indirectly coupled to asynchronous callback execution and rendering-driven timing behavior. Conventional deterministic execution assumptions, including stable execution ordering and reproducible scheduling behavior, become harder to maintain when simulation progression depends on event-loop scheduling, rendering cadence, and asynchronous Worker coordination.

\subsection{Simulation--Rendering Coupling}

Many browser-based interactive systems advance simulation state directly from rendering callbacks such as \url{requestAnimationFrame}. Under this organization, simulation progression becomes partially coupled to rendering cadence and browser timing behavior.

Because rendering callbacks execute asynchronously relative to Worker updates, rendering stages may sample simulation state while partition-local updates are still incomplete. Different rendering and callback timing conditions may therefore expose different intermediate states across executions, causing simulation progression to depend on runtime timing behavior rather than solely on logical simulation inputs.

\vspace{0.5em}
\noindent
\textbf{Limitation.}
Coupling simulation progression directly to rendering behavior makes execution results more sensitive to runtime timing variability and asynchronous callback scheduling.

\subsection{Unifying Perspective and Research Gap}

Prior approaches use different mechanisms to reduce nondeterministic execution, including synchronization control, deterministic scheduling, causal ordering, and execution replay. Despite these differences, most assume that runtime ordering remains sufficiently stable to be coordinated, observed, or reproduced across executions.

This assumption becomes harder to maintain in browser execution environments. Browser applications progress through asynchronously dispatched callbacks originating from multiple runtime subsystems, including rendering, user interaction, networking, and Worker communication. Execution ordering may therefore vary across runs even when application logic and input conditions remain unchanged.

In particular, rendering progression and simulation progression are often only weakly coordinated. Callback dispatch timing may vary with rendering load, event-loop activity, browser implementation behavior, and external event delivery. Consequently, preserving identical simulation results becomes difficult when logical state progression depends directly on asynchronous callback ordering.

\subsection{Positioning of This Work}

In contrast to prior approaches, this work defines simulation progression through logical-time-driven state updates:
\begin{equation*}
S_{k+1} = F(S_k, t_k)
\end{equation*}
where each simulation step is computed from the current logical state and logical time.

Under this organization, runtime callback ordering no longer directly determines when updated simulation states become externally observable. Rendering cadence, event-loop timing, and Worker scheduling may still affect execution latency, but they do not alter committed logical-state progression across executions.

\vspace{0.5em}
\noindent
\textbf{Key Distinction.}
Instead of controlling execution order directly, the proposed model constrains how simulation states are updated during each logical step, so that execution results remain consistent across different scheduling conditions. Under this organization, logical simulation progression becomes less sensitive to callback dispatch timing and scheduling differences.
\section{RepliCore Execution Model}

\subsection{Problem Formulation}

Large-scale browser-oriented simulations commonly execute over asynchronous runtime environments in which simulation states may become externally visible at different physical times across executions due to independently progressing Workers, rendering callbacks, and event-loop scheduling. 

Independent Workers may progress at different rates, causing some partition updates to complete earlier than others. If partially updated states become externally visible before logical-state progression reaches the next step boundary, subsequent simulation updates, rendering callbacks, or asynchronous event handlers may consume inconsistent logical-time state information. Repeated executions may therefore construct different logical update sequences even under identical simulation inputs.

For large-scale simulation systems, this behavior introduces several practical issues, including inconsistent results between sequential and parallel execution, reduced reproducibility, and unstable behavior across different parallel configurations.

To address these issues, the execution model should satisfy the following requirements:
\begin{itemize}
\item \textbf{Reproducible execution:} repeated runs initialized from the same state should produce identical committed simulation states;
\item \textbf{Sequential-parallel consistency:} parallel execution should preserve the same logical state progression as sequential execution;
\item \textbf{Scheduling tolerance:} Worker scheduling and callback timing variations should not alter committed simulation states;
\item \textbf{Rendering separation:} rendering stages should observe only completed logical simulation states;
\item \textbf{Parallel scalability:} increasing Worker parallelism should not introduce additional state divergence.
\end{itemize}

These requirements motivate an execution organization that controls when updated simulation states become externally visible during logical-time progression.


\subsection{Logical-Time Visibility Boundaries}

Simulation progression cannot rely directly on physical execution timing because browser Workers and rendering callbacks advance independently under asynchronous runtime scheduling. RepliCore therefore advances logical-state progression through logical-time commitment boundaries shared across asynchronous runtime stages. Given a fixed simulation step size $\Delta t$, logical time advances discretely as

\begin{equation}
\label{eq:logical_time}
t_k = k \Delta t, \quad k \in \mathbb{N}
\end{equation}

State updates are performed only at explicit simulation-step boundaries $t_k$.

Logical time defines explicit step boundaries shared across Workers, state updates, and rendering stages. Updated simulation states become externally visible only after logical-state progression reaches a stable boundary at the current logical step. All Workers, rendering stages, and subsequent simulation updates therefore observe the same committed logical-time state rather than partially completed intermediate updates.

Variations in Worker execution speed or callback scheduling may affect runtime latency, but committed simulation trajectories remain consistent across logical-time boundaries.


\subsection{Deterministic State Updates}

Consistent logical-state progression requires all logical-time updates to derive from the same committed simulation-state snapshot.

Let $S_k$ denote the committed simulation state at logical time $t_k$.
At each logical-time advancement, the runtime derives the next committed state through deterministic state propagation over the logical-time interval $\Delta t$:
\[
S(t_{k+1}) = \Phi(S(t_k), \Delta t)
\]
where $\Phi$ denotes the state-propagation procedure executed during the current logical-time interval.
No Worker may observe or access $S_{k+1}$ before state construction for the current logical-time step completes.

Within each logical-time step, all state updates operate on the same committed logical-state snapshot $S_k$. The resulting logical-time advancement therefore derives from the same committed-state snapshot rather than from runtime-dependent Worker scheduling or message-delivery timing.

This update model prevents partially completed updates from becoming externally visible before logical-state progression stabilizes. Consequently, runtime timing differences cannot change which logical state snapshot is consumed by subsequent updates or rendering stages.


\subsection{Parallel Partitioned Execution}

State coordination under asynchronous runtime execution requires partition-local updates to remain isolated before the next committed state is published. To support this, the global simulation state is divided into disjoint partitions:

\begin{equation}
S_k = \bigsqcup_{i=1}^{n} S_k^{(i)}
\end{equation}

Each partition is executed by an independent Worker and performs local state updates on partition-local simulation data during the current logical-time step:

\begin{equation}
S_{k+1}^{(i)} = F_i(S_k^{(i)})
\end{equation}

After partition-local updates for the current logical-time step have been collected, the runtime combines these resulting partition states to construct the next global state:

\begin{equation}
S_{k+1} = \mathcal{A} \left( \{ S_{k+1}^{(i)} \}_{i=1}^{n} \right)
\end{equation}

where $\mathcal{A}$ represents the deterministic state-construction procedure applied at the current logical-time boundary.

During each logical-time step, local computations remain partition-scoped, while updated states become externally visible only after the current logical-time step reaches a stable visibility boundary. This execution structure reduces dependence on Worker scheduling order and helps preserve consistent committed simulation states across different parallel configurations.


\subsection{Execution Requirements}

Maintaining consistent logical-state progression under asynchronous runtime execution requires several visibility and execution constraints:

\begin{itemize}
\item \textbf{R1: Deterministic partition execution} — partition-local updates must operate consistently for the same logical input state;
\item \textbf{R2: Isolated partition writes} — partition-local updates cannot expose intermediate writes before committed-state construction completes;
\item \textbf{R3: Stable state visibility} — partition results should be integrated through a runtime-consistent state-construction procedure before becoming externally visible
\item \textbf{R4: Logical-time advancement} — the next committed logical-time state cannot become externally visible until partition-local updates associated with the current step become available.
\end{itemize}

These conditions define the execution constraints required to preserve logical-time state consistency under asynchronous runtime execution.


\subsection{Parallel Consistency and Schedule Independence}

During execution, controlled logical-state visibility preserves the same logical simulation-state progression across sequential and parallel execution:

\begin{equation}
S_{k+1}^{\text{parallel}} 
= \mathcal{A} \left( \{ F_i(S_k^{(i)}) \} \right) 
= 
F(S_k)
=
S_{k+1}^{\text{sequential}}
\end{equation}

This execution model allows parallel execution to produce the same committed simulation states as sequential execution while reducing dependence on runtime scheduling and asynchronous timing behavior.

When all partition-local updates operate on the same logical input snapshot and state publication occurs only after committed-state construction, Worker scheduling variations do not alter committed simulation states.

Runtime scheduling may still influence execution latency and resource utilization, but browser-level execution variability does not change logical simulation results across executions.


\subsection{Rendering Decoupling}

Rendering behavior is decoupled from logical simulation-state progression under asynchronous browser runtime execution:

\begin{equation}
R_k = G(S_k)
\end{equation}

where $G$ represents the rendering procedure applied to the current simulation state.

During logical-time progression, rendering consumes only committed logical simulation states produced after committed-state construction. Rendering accesses only completed logical states after state construction finishes. Since rendering callbacks execute asynchronously relative to Worker execution, this separation prevents rendering stages from sampling partially updated simulation states during logical-step progression.

Rendering does not participate in simulation state updates and therefore cannot alter logical simulation state progression.

This separation establishes an explicit boundary between simulation updates and rendering operations, reducing interference between render-frame timing, asynchronous callback execution, and state publication.


\subsection{Implementation Implications}

The execution model targets browser-oriented simulation systems in which state progression must remain consistent despite asynchronous Worker execution, rendering activity, and shared runtime services.

Implementations must coordinate partition-local state progression, committed-state advancement, and cross-runtime state consistency while limiting unintended shared-state exposure during execution.

Logical-time visibility control introduces additional runtime overhead through state buffering, inter-Worker message exchange, and state-construction management.

Execution consistency further depends on stable partition assignment and consistent handling of cross-partition interactions and state publication ordering.

Although these coordination constraints may reduce peak throughput under smaller workloads, they help maintain reproducible simulation behavior under asynchronous browser runtime scheduling.

Reproducible committed-state progression further assumes deterministic partition-local update behavior, stable partition assignment, and deterministic state-construction ordering under a fixed browser runtime and numerical execution environment. The current implementation does not attempt to enforce cross-engine floating-point equivalence across heterogeneous JavaScript runtimes.
\section{RepliCore Architecture and Implementation}

RepliCore targets browser-based parallel simulations executed over asynchronous runtime environments composed of Workers, event loops, asynchronous message passing, and rendering callbacks.

Because these runtime components progress independently, partially updated simulation states may become visible before a logical step reaches a consistent state.

To preserve reproducible simulation behavior under these runtime conditions, RepliCore organizes execution into repeated logical-time steps coordinated by a Master thread. Workers compute partition-local updates independently and return local results through asynchronous message passing.

The architecture consists of a Master thread, partitioned Workers, a runtime coordinator, and a deterministic state-construction stage.


\subsection{Visibility Coordination}

RepliCore coordinates when updated simulation states become externally observable across asynchronous browser runtime stages.

\begin{itemize}
\item partition-local execution over isolated Worker state
\item committed-state coordination across asynchronous execution stages
\item deterministic construction of committed simulation state
\item rendering decoupling from intermediate runtime updates
\end{itemize}

Together, these mechanisms prevent partially updated states from becoming externally visible during logical-state progression.

Workers execute updates independently on partition-local state. Intermediate results remain locally buffered until all partition updates for the current logical step are available. The runtime then deterministically constructs the next global state and publishes it to subsequent execution and rendering stages.

This organization prevents partially completed updates and asynchronous Worker completion order from influencing global state construction.


\subsection{Overall Architecture}

RepliCore uses a Master--Worker runtime structure implemented over browser Workers and asynchronous message passing.

The runtime consists of the following components:
\begin{itemize}
\item \textbf{Master:} advances logical time and coordinates execution stages
\item \textbf{Partition Scheduler:} decomposes global state into isolated partitions
\item \textbf{Workers:} execute local state updates in parallel
\item \textbf{Runtime Coordinator:} coordinates committed global state across asynchronous Worker execution
\item \textbf{State Constructor:} constructs the next committed state from partition-local updates
\item \textbf{Renderer:} consumes committed states for visualization
\end{itemize}

The Master dispatches partition updates, tracks partition-local update completion, coordinates state construction, and advances logical-time steps. Workers execute partition-local updates concurrently without directly modifying shared global state. After partition-local updates associated with a logical-time boundary have been collected, the State Constructor deterministically produces the next committed state.

\begin{figure}[t]
\centering
\resizebox{\columnwidth}{!}{
\begin{tikzpicture}[
    node distance=1.1cm,
    box/.style={
        rectangle,
        draw,
        thick,
        minimum width=2.5cm,
        minimum height=0.75cm,
        align=center,
        rounded corners=2pt
    },
    worker/.style={
        rectangle,
        draw,
        thick,
        minimum width=1.6cm,
        minimum height=0.65cm,
        align=center
    },
    arrow/.style={-{Stealth[length=2.5mm]}, thick}
]

\node[box] (sk) {$S_k$};

\node[box, below=of sk] (part) {Partitioning};

\node[worker, below=1.4cm of part, xshift=-3.2cm] (w1) {Worker};
\node[worker, right=1.6cm of w1] (w2) {Worker};
\node[worker, right=1.6cm of w2] (w3) {Worker};

\node[
    draw,
    dashed,
    rounded corners,
    fit=(w1)(w2)(w3),
    inner sep=0.5cm,
    label=above:{Asynchronous Workers}
] {};

\node at ($(w2)!0.5!(w3)$) [above=-0.35cm] {$\cdots$};

\node[box, below=1.4cm of w2] (commit)
{State Publication};

\node[box, below=of commit] (agg)
{State Construction};

\node[box, below=of agg] (sk1)
{$S_{k+1}$};

\node[box, below=of sk1] (render)
{Rendering};

\draw[arrow] (sk) -- (part);

\draw[arrow] (part.south) -- ++(0,-0.35) -| (w1.north);
\draw[arrow] (part.south) -- ++(0,-0.35) -| (w2.north);
\draw[arrow] (part.south) -- ++(0,-0.35) -| (w3.north);

\draw[arrow] (w1.south) -- ++(0,-0.35) -| (commit.north);
\draw[arrow] (w2.south) -- (commit.north);
\draw[arrow] (w3.south) -- ++(0,-0.35) -| (commit.north);

\draw[arrow] (commit) -- (agg);
\draw[arrow] (agg) -- (sk1);
\draw[arrow] (sk1) -- (render);

\end{tikzpicture}
}
\caption{
Runtime organization of RepliCore.
Workers execute partition-local updates independently before the runtime constructs the next committed state.
}
\label{fig:architecture}
\end{figure}
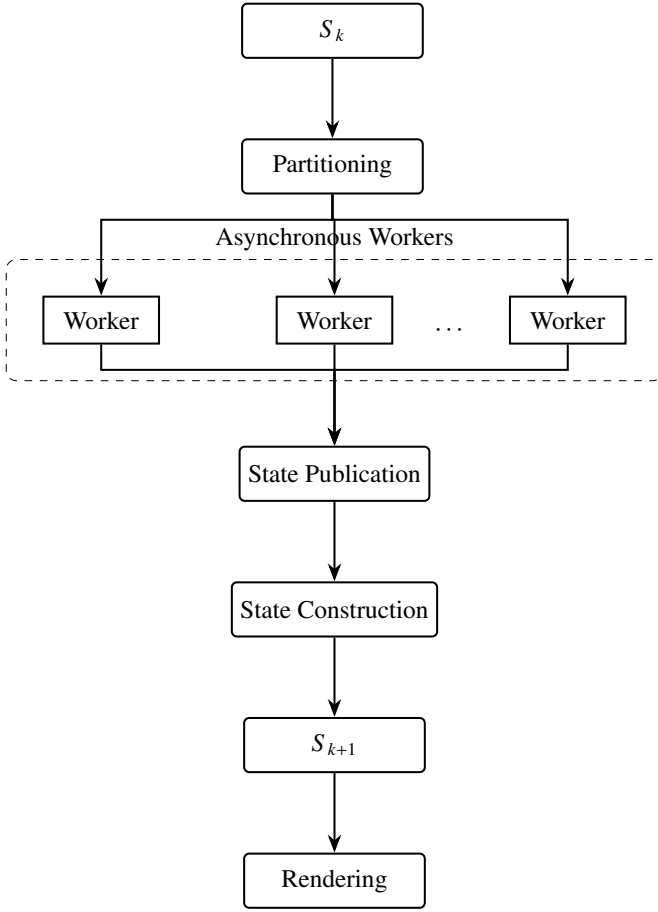

Fig.~\ref{fig:architecture} illustrates the overall runtime organization of RepliCore. At each logical-time boundary, the runtime advances observable simulation-state progression by isolating partition-local updates, controlling committed-state across asynchronous Worker execution, and resolving the next externally observable logical-time state.

Under this runtime structure, committed state construction remains independent of asynchronous Worker completion order.

\paragraph{Runtime organization}

The execution flow shown in Fig.~\ref{fig:architecture} proceeds as follows. The Master partitions the committed state and dispatches partition-local updates to Workers. Workers execute independently during a logical-time step and return local update results to the coordination layer.

The coordination layer delays state publication until partition-local updates associated with the current logical-time boundary have been collected. The coordination layer then deterministically constructs the next committed state from partition-local update results.

In this organization, all Workers derive updates from the same committed input state $S_k$, while $S_{k+1}$ becomes externally visible only after logical-time coordination and state construction completes.


\subsection{Execution Flow}

Observable simulation-state progression advances through repeated committed-state transitions under asynchronous execution.

Execution proceeds in discrete logical-time steps:

\begin{enumerate}
\item Read the committed global state $S_k$ at logical time $t_k$
\item Partition $S_k$ into disjoint subsets $S_k^{(i)}$
\item Dispatch partition-local updates to Workers
\item Compute local update results $S_{k+1}^{(i)}$
\item   Collect partition-local updates for committed-state construction
\item Merge local results into the next global state $S_{k+1}$
\item Render output from the committed state
\end{enumerate}

To preserve consistent logical-time state visibility during asynchronous execution, the runtime follows several coordination rules:

\begin{itemize}
\item \textbf{Single-input execution:} all Worker computations within a logical step depend only on $S_k$
\item \textbf{Visibility consistency:} intermediate partition updates remain non-observable before committed-state construction completes
\item \textbf{Single committed state:} the next global state becomes externally visible only after deterministic committed-state construction
\end{itemize}

These rules ensure that all Workers execute from the same committed logical-time snapshot and that state construction remains independent of asynchronous Worker completion order.


\subsection{Partition Isolation}

Observable-state consistency becomes difficult to preserve when asynchronous runtime execution exposes shared mutable state across independently progressing Workers. To reduce unintended interactions across asynchronous worker execution, RepliCore isolates state updates at the partition level.

For disjoint partitions:
\begin{equation*}
\forall i \neq j,\quad
\text{Read}(S_k^{(i)}) \cap \text{Write}(S_k^{(j)}) = \emptyset
\end{equation*}

Under partition-local execution, Workers read only partition-local data during a logical-time step, and updates are produced independently before committed-state coordination.

\textbf{Implementation.}
\begin{itemize}
\item Workers operate on isolated partition-local state
\item partition data are exchanged through message passing rather than shared writable memory
\item direct concurrent writes across Workers are disallowed
\item cross-partition interaction is handled explicitly during committed-state coordination and construction
\end{itemize}

Partition isolation reduces unintended execution dependencies and helps maintain consistent state updates across asynchronous execution schedules.


\subsection{Visibility Coordination Mechanism}

Asynchronous browser runtime execution may expose partially updated simulation states at inconsistent logical-step boundaries.

RepliCore therefore constrains when updated states become externally visible at logical-step boundaries so that subsequent execution and rendering stages observe a consistent global state snapshot.

The current implementation uses a step coordination stage that delays state publication until all partition-local updates for the current logical step become available.

\begin{equation*}
S_{k+1} \text{ becomes externally visible}
\;\iff\;
\forall i,\; S_{k+1}^{(i)} \text{ are available}
\end{equation*}

The coordination mechanism prevents partially updated states from becoming globally visible during parallel execution. Intermediate partition updates remain non-observable until committed-state resolution completes. Without this constraint, different Workers or rendering stages may observe inconsistent committed-state snapshots during execution, causing runtime timing behavior to influence subsequent state updates.

The current implementation realizes committed-state coordination through a coordination stage between logical-time boundaries. The coordination layer delays committed-state exposure until partition-local updates associated with the current logical-time boundary have been collected for deterministic state construction.

Subsequent execution stages therefore observe the same committed logical-state snapshot.


\subsection{Deterministic State Construction}

Consistent committed-state progression further depends on stable committed-state construction under asynchronous update completion. If local updates are merged in different orders, the resulting global state may vary across executions.

For a merge operation $\mathcal{A}$:
\begin{equation*}
\mathcal{A}(x,y) \neq \mathcal{A}(y,x)
\end{equation*}
can produce different outcomes under asynchronous execution schedules.

To maintain consistent state construction, committed-state construction can be organized either through order-independent operations (e.g., commutative and associative reductions) or through an explicitly defined merge order.

RepliCore uses deterministic committed-state construction primarily to preserve consistent global-state construction across repeated executions under asynchronous runtime scheduling. This approach simplifies deterministic state construction and provides more stable floating-point accumulation behavior across repeated executions.

Committed state construction therefore depends on the logical-time update sequence rather than asynchronous Worker completion order.


\subsection{Rendering Decoupling}

Observable rendering behavior may diverge across executions when rendering callbacks sample partially updated simulation states under browser execution.

In browser-based simulation systems, rendering callbacks and simulation updates often execute at different frequencies. In RepliCore, rendering is separated from logical-time state updates to prevent rendering callback timing from influencing simulation results.

The rendering process is defined as:
\begin{equation*}
R_k = G(S_k), \quad G \notin F
\end{equation*}
where rendering depends on the committed simulation state but does not participate in state update computation.

Rendering observes only committed logical-time snapshots produced after coordinated state commitment. Variations in rendering frequency or display timing do not affect logical-time state visibility or state construction.

This separation prevents rendering callback timing from affecting inconsistent state sampling and logical-time visibility progression across browser runtime environments.

\subsection{Runtime Coordination Summary}

RepliCore organizes browser-based parallel simulation around logical-step state visibility control under asynchronous runtime execution.

Partition-local updates remain non-observable during execution, while runtime-wide coordination constructs a single committed logical-time snapshot for subsequent execution and rendering.

This organization reduces observable-state divergence introduced by asynchronous Worker scheduling, message delivery timing, and rendering callbacks.
\section{Experimental Evaluation}

The evaluation examines whether RepliCore preserves reproducible simulation execution under asynchronous browser runtime conditions involving concurrent rendering, event-loop scheduling, and Worker execution variability.

The experiments evaluate:
\begin{itemize}
\item reproducibility under varying Worker scheduling and rendering conditions;
\item execution divergence after relaxing logical progression constraints;
\item scalability under coordinated parallel execution.
\end{itemize}


\subsection{Experimental Setup}

The experiments evaluate simulation execution under browser runtime conditions involving concurrent rendering and independently progressing Workers.

The evaluated workload represents a browser-oriented large-scale entity simulation with continuously evolving state trajectories under asynchronous browser execution. The implementation uses a satellite-style propagation workload to emulate long-running state evolution across partitioned Workers. Each Worker independently updates partition-local satellite states through repeated orbital-state propagation, while browser-side rendering callbacks concurrently visualize evolving simulation states during asynchronous event-loop execution.

Additional browser-side rendering load was introduced through continuously active rendering callbacks, frame-driven visual-state updates, and asynchronous browser event-loop activity during simulation execution.

\paragraph{Workload characteristics}

The evaluated workload represents a browser-oriented parallel simulation with continuously evolving entity states and repeated cross-step floating-point updates. During each logical-time step, Workers update partition-local entities independently, after which local updates are merged into a deterministic global state visible to subsequent simulation and rendering stages.

Because entity states evolve incrementally over many logical-time steps, small execution differences may accumulate over time, making the workload suitable for evaluating reproducibility under asynchronous execution.

\textbf{Platform configuration:}
\begin{itemize}
    \item CPU: Intel Ultra 9 (16 cores)
    \item Memory: 32GB
    \item Runtime: Chrome (V8)
\end{itemize}

\textbf{Simulation parameters:}
\begin{itemize}
    \item Logical time step: $\Delta t = 1$
    \item Total logical steps: 10,000
    \item Simulation scale: $10^4$ to $2\times10^6$ entities
    \item Parallel execution: up to 16 Workers
\end{itemize}

Each configuration was evaluated across repeated independent executions. Coordinated configurations consistently produced bitwise-identical outputs and identical trajectory hashes, with timing variance remaining below 1\%.


\subsection{Execution Reproducibility under Runtime Variability}

This experiment evaluates whether simulation execution remains reproducible under varying runtime scheduling and rendering conditions.

\textbf{Method.}

The same simulation workload was executed under different runtime conditions, including:
\begin{itemize}
\item varying Worker configurations (1 / 2 / 4 / 8 / 16 Workers),
\item varying rendering frequencies (disabled / 60 FPS / 30 FPS / 10 FPS),
\item concurrent browser-side rendering activity,
\item asynchronous timer callbacks executing during simulation progression,
\item artificial Worker timing perturbation introduced through randomized execution delays.
\end{itemize}

To emulate unstable browser scheduling conditions, randomized timing delays were dynamically injected into Worker execution during selected runs. The perturbation delays varied independently across Workers throughout execution in order to emulate unstable runtime scheduling conditions.

\textbf{Metrics.}
Execution reproducibility across repeated runs was evaluated using:
\begin{itemize}
    \item bitwise-identical simulation states across runs,
    \item identical SHA-256 trajectory hashes across runs.
\end{itemize}

Trajectory hashes were computed from periodically sampled simulation states during execution.

\textbf{Results.}

\begin{table}[t]
\centering
\small
\caption{Execution reproducibility across varying asynchronous Worker configurations}
\label{tab:workerscal}
\begin{tabular}{c c c}
\toprule
Workers & Bitwise-Identical States & Trajectory Hashes \\
\midrule
1 & Yes & Identical \\
2 & Yes & Identical \\
4 & Yes & Identical \\
8 & Yes & Identical \\
16 & Yes & Identical \\
\bottomrule
\end{tabular}
\end{table}

\begin{table}[t]
\centering
\small
\caption{Execution reproducibility under varying rendering frequencies}
\label{tab:render}
\begin{tabular}{c c c}
\toprule
Render Frequency & Bitwise-Identical States & Trajectory Hashes \\
\midrule
Off & Yes & Identical \\
60 & Yes & Identical \\
30 & Yes & Identical \\
10 & Yes & Identical \\
\bottomrule
\end{tabular}
\end{table}

As shown in Table~\ref{tab:workerscal} and Table~\ref{tab:render}, all evaluated configurations produced identical trajectory hashes and bitwise-identical simulation states across repeated executions.

The Worker-configuration experiments were conducted using 1, 2, 4, 8, and 16 Workers in order to evaluate reproducibility under different parallel Worker execution configurations. No cross-run state divergence was observed under any evaluated Worker configuration.

Rendering-frequency experiments were performed under the 16-Worker configuration using disabled rendering as well as 60 FPS, 30 FPS, and 10 FPS rendering rates. All configurations produced bitwise-identical states across runs. All evaluated configurations produced identical outputs under the tested Worker perturbations, asynchronous event-loop callbacks, and concurrent rendering activity.

Across all evaluated configurations, repeated executions produced identical outputs despite variations in Worker scheduling, rendering frequency, and injected timing perturbations.


\subsection{Coordination Relaxation Analysis}

This experiment evaluates how relaxing logical-time visibility coordination affects logical-state progression under asynchronous runtime execution.

\textbf{Method.}    
The workload was executed under several coordination configurations:
\begin{itemize}
\item \textbf{Baseline:} coordinated global-state commitment enabled across logical-time progression
\item \textbf{Commitment-Decoupled:} state publication allowed before all Worker updates for the current step completed
\item \textbf{Render-Coupled:} logical-time advancement additionally driven by browser rendering callbacks
\end{itemize}

\textbf{Results.}

State divergence was measured as the accumulated numerical difference between corresponding entity states relative to the coordinated committed-state baseline:

\[
D_k = \sum_{i=1}^N \left\| s_{k,i}^{\text{test}} - s_{k,i}^{\text{baseline}} \right\|
\]

where $\|\cdot\|$ denotes the Euclidean distance between corresponding entity-state vectors. Each entity-state vector contains normalized position and velocity components represented in simulation-space units.

This metric measures how timing-dependent state differences accumulate after logical-step consistency constraints are relaxed during execution.

Table~\ref{tab:baseline} presents the comparison.
Mean and peak divergence values were computed over sampled logical-time states across repeated executions relative to the coordinated committed-state baseline.

\begin{table}[t]
\centering
\small
\caption{State divergence after relaxing coordinated state publication}
\label{tab:baseline}
\begin{tabular}{l c c}
\toprule
Configuration & Mean Div. & Peak Div. \\
\midrule
RepliCore & 0 & 0 \\
Early Publication & $1.7\times10^6$ & $6.4\times10^6$ \\
Render-Coupled & $2.6\times10^6$ & $6.4\times10^6$ \\
\bottomrule
\end{tabular}
\end{table}

The results suggest that allowing partially updated states to become available during execution can alter subsequent state evolution under asynchronous runtime conditions.


\subsection{Scalability under Coordinated Parallel Execution}

This experiment evaluates how coordinated committed-state progression behaves under increasing browser-oriented simulation scale.

\textbf{Method.}

Execution time and browser responsiveness were evaluated under increasing simulation scales for both serial and Worker-based parallel execution.

\textbf{Results.}

\begin{table*}[t]
\centering
\small
\caption{Runtime scalability under coordinated execution}
\label{tab:scalability}
\begin{tabular}{c c S S S S}
\toprule
Scale & Mode & {Time (ms)} & {Throughput (entities/s)} & {Cost (ms/entity)} & {Speedup} \\
\midrule

10K & Serial   & 72.25   & 138350 & 0.00723 & {} \\
10K & Parallel & 257.86  & 38780  & 0.02579 & 0.28 \\

100K & Serial   & 753.55  & 132730 & 0.00754 & {} \\
100K & Parallel & 2071.85 & 48266  & 0.02072 & 0.36 \\

500K & Serial   & 24500   & 20408  & 0.04900 & {} \\
500K & Parallel & 15725.11 & 32346 & 0.03146 & 1.56 \\

1M & Serial   & 26739.44 & 37388 & 0.02674 & {} \\
1M & Parallel & 22380.50 & 44682 & 0.02238 & 1.19 \\

2M & Serial   & 56563.16 & 35365 & 0.02828 & {} \\
2M & Parallel & 47605.63 & 42018 & 0.02380 & 1.19 \\

\bottomrule
\end{tabular}
\end{table*}

\begin{table}[t]
\centering
\small
\caption{Observed browser responsiveness during coordinated logical-state progression}
\label{tab:responsiveness}
\begin{tabular}{c c c}
\toprule
Scale & Serial & Parallel \\
\midrule
100K & Responsive & Responsive \\
500K & Intermittent blocking & Responsive \\
1M & Unresponsive & Responsive \\
2M & Unresponsive & Responsive \\
\bottomrule
\end{tabular}
\end{table}

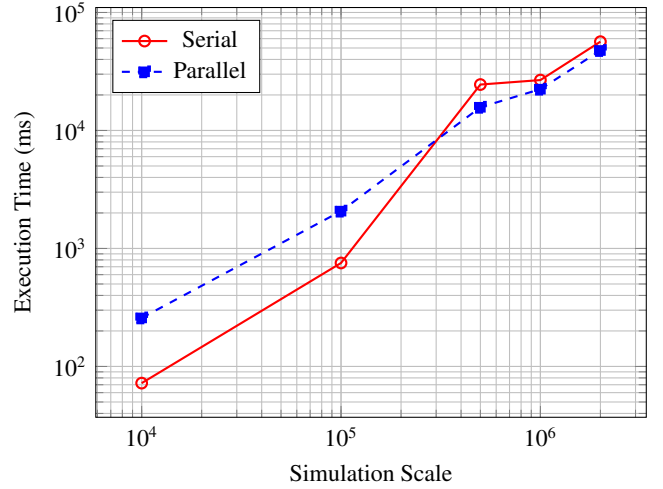
\begin{figure}[t]
\centering
\begin{tikzpicture}
\begin{axis}[
    width=\linewidth,
    height=7cm,
    xlabel={Simulation Scale},
    ylabel={Execution Time (ms)},
    xmode=log,
    ymode=log,
    grid=both,
    legend pos=north west,
    tick label style={font=\small},
    label style={font=\small},
    legend style={font=\small},
]

\addplot[thick, red, mark=o] coordinates {
(1e4,72.25)
(1e5,753.55)
(5e5,24500)
(1e6,26739.44)
(2e6,56563.16)
};
\addlegendentry{Serial}

\addplot[thick, dashed, blue, mark=square*] coordinates {
(1e4,257.86)
(1e5,2071.85)
(5e5,15725.11)
(1e6,22380.50)
(2e6,47605.63)
};
\addlegendentry{Parallel}

\end{axis}
\end{tikzpicture}
\caption{Execution time scaling under serial and parallel execution configurations. Serial execution becomes unresponsive at large scales. Error bars are omitted due to negligible variance ($<1\%$).}
\label{fig:time_scale}
\end{figure}

As shown in Table~\ref{tab:scalability} and Fig.~\ref{fig:time_scale}, parallel execution introduces additional coordination and state-aggregation overhead at smaller scales, resulting in lower throughput than direct serial execution.

At larger scales, the additional coordination overhead is offset by increased parallel computation, allowing parallel execution to outperform serial execution. Under large-scale workloads, the parallel configuration reduces main-thread execution pressure while preserving reproducible simulation progression across asynchronous Worker execution.

Browser responsiveness was also evaluated during execution. During large-scale serial execution, simulation updates executed directly on the browser main thread and increasingly interfered with rendering callbacks and event-loop responsiveness. At scales beyond 500K entities, rendering updates and user-interface interaction became intermittently blocked, as shown in Table~\ref{tab:responsiveness}.

Under Worker-based execution, simulation computation progressed independently from rendering and event-loop execution. Rendering callbacks and browser-side interaction therefore remained responsive despite increasing simulation scale.

These results indicate that coordinated parallel execution introduces measurable coordination overhead while remaining practical for large-scale browser-oriented simulation workloads.


\subsection{Summary of Empirical Findings}

The current evaluation focuses primarily on runtime coordination behavior and execution reproducibility under asynchronous browser execution. More complex simulation workloads with dynamically changing interaction topology remain important future evaluation targets.

The evaluation shows that:

\begin{itemize}
\item coordinated step-level state publication preserves reproducible simulation execution across varying Worker and rendering configurations;
\item relaxing step-level coordination was associated with measurable execution divergence
\item Worker-based execution preserves browser responsiveness under increasing simulation scale despite runtime coordination overhead.
\end{itemize}
\section{Discussion}

\subsection{Deterministic Execution and Runtime Reproducibility}

RepliCore organizes simulation execution around coordinated state publication under asynchronous browser runtime execution.

Many deterministic execution systems implicitly assume that execution ordering and state visibility can be constrained or reproduced consistently across executions. This assumption becomes increasingly fragile in browser runtimes, where rendering callbacks, event-loop activities, asynchronous message delivery, and independently progressing Workers may observe simulation states at different physical times.

RepliCore maintains reproducible execution by controlling when updated states become globally visible during simulation progression. Workers may execute at different speeds, while rendering callbacks and browser event handling continue asynchronously during execution. The runtime publishes state updates only after the current logical step has been resolved, reducing dependence on Worker completion timing and callback interleavings.

The evaluation indicates that this organization preserves stable simulation results across varying rendering conditions and Worker scheduling behavior within a fixed browser runtime environment.


\subsection{Trade-offs and Structural Constraints}

Maintaining coordinated state publication under asynchronous runtime execution introduces several structural trade-offs.

Coordinated logical-time commitment reduces the ability of Workers to expose intermediate simulation states independently. Under highly imbalanced workloads, faster Workers may remain temporarily stalled before committed-state visibility can advance globally. As the number of Workers increases, overall progress may become increasingly influenced by the slowest partition update within each logical step.

Deterministic execution requires state updates to depend only on explicitly coordinated committed-state inputs. Execution patterns that rely on shared mutable state or runtime-dependent interactions are therefore more difficult to support directly.

The execution model also assumes that simulation updates can be decomposed into relatively isolated partitions. Systems with dense cross-partition interaction or rapidly changing interaction topology may require substantially more runtime coordination and data exchange between logical-time steps.

Rendering decoupling further separates simulation updates from frame timing. This improves reproducibility under unstable rendering conditions, but it also reduces the ability of interactive applications to directly couple simulation updates with visual refresh behavior.

In practice, the proposed organization trades part of the flexibility of fully asynchronous execution for more stable and reproducible simulation behavior.


\subsection{Applicability}

RepliCore is most suitable for simulation workloads in which entities can be partitioned into relatively stable execution regions and updated independently during a logical step. Typical examples include browser-based entity simulation, digital twin visualization, and interactive large-scale simulation workloads.

The model becomes less effective when simulation behavior depends heavily on rapidly changing cross-partition interaction or fine-grained shared-state communication, since these conditions increase coordination and data-exchange overhead between Workers.


\subsection{Limitations and Future Work}

A central limitation of the proposed execution model is the requirement that state visibility advances only after each logical step reaches a globally consistent state. As interaction density increases, runtime coordination and communication overhead may grow substantially, particularly in systems with fine-grained cross-partition dependencies or dynamically changing interaction topology. The current implementation targets reproducible execution within a fixed browser engine and numerical runtime environment. RepliCore does not attempt to guarantee cross-engine equivalence across heterogeneous JavaScript implementations or hardware-dependent floating-point behavior.

An important next step is exploring whether parts of the visibility-coordination process can be relaxed without reintroducing schedule-dependent divergence. This includes partially coordinated execution, adaptive coordination boundaries, and more efficient coordination mechanisms for dynamically coupled simulation workloads.

\section{Conclusion}

The results suggest that reproducible parallel simulation in browser runtime environments can be achieved through controlled state publication under asynchronous Worker execution and rendering activity.

The evaluation shows stable and reproducible execution across varying Worker configurations, rendering frequencies, and injected runtime perturbations within a fixed browser runtime environment.

Although coordinated execution introduces additional coordination overhead, the approach remains practical for large-scale browser-oriented simulation workloads while preserving responsive rendering behavior.

Future work will explore adaptive coordination mechanisms, distributed multi-node execution, and execution models for more dynamically coupled simulation workloads.

\section*{Acknowledgment}
During the preparation of this work the author used ChatGPT in order to improve the clarity and readability of this manuscript. After using this tool, the author reviewed and edited the content as needed and takes full responsibility for the content of the publication.

\bibliographystyle{elsarticle-num}
\bibliography{refs}

\end{document}